\documentclass[%
 reprint,
preprint,
 amsmath,amssymb,
aps,
onecolumn
]{revtex4-1}

\usepackage{graphicx}
\usepackage{dcolumn}
\usepackage{bm}
\usepackage{color}
\usepackage{ulem}

\begin{document}

\preprint{APS/123-QED}

\title{\textcolor{black}{Low Curie-temperature ferromagnetic phase in SmPt$_{2}$Cd$_{20}$ possibly accompanied by strong quantum fluctuations}\\}

\author{Akira Yamada}\email{yamada-akira1@ed.tmu.ac.jp}
\author{Shota Oike, Ryuji Higashinaka, Tatsuma D. Matsuda}
\author{Yuji Aoki}\email{aoki@tmu.ac.jp}
\affiliation{%
Department of Physics, Tokyo Metropolitan University, Hachioji, Tokyo 192-0397, Japan\\
}%

\date{\today}

\begin{abstract}
Electrical resistivity, magnetization and specific heat have been measured for single crystals of SmPt$_{2}$Cd$_{20}$.
It has been found that SmPt$_{2}$Cd$_{20}$ exhibits a ferromagnetic (FM) transition at $T_{\rm C} = 0.64$ K, the lowest among cubic compounds.
Specific heat divided by temperature increases with decreasing temperature even below $T_{\rm C}$ and attains 4.5 J/mol K$^{2}$ at 0.26 K, implying substantial magnetic quantum fluctuations.
An analysis of the magnetic entropy suggests the crystalline-electric-field splitting of the Sm $J = 5/2$ multiplet with a $\Gamma_{7}$ doublet ground state and a $\Gamma_8$ quartet excited state (the excitation energy of $\sim30$ K). 
The electrical resistivity shows a power-law behavior with \textcolor{black}{$T^{0.74}$} below 2 K without showing any noticeable anomaly at $T_{\rm C}$.
SmPt$_{2}$Cd$_{20}$ is regarded as a rare cubic system that is located in the vicinity of a FM quantum critical point.
\end{abstract}

\pacs{Valid PACS appear here}
\maketitle


Quantum fluctuations of magnetism are expected to dominate at around $T = 0$ and to cause unconventional superconductivity and other anomalous electronic states in metals.
By tuning some nonthermal control parameters such as pressure ($P$), doping or magnetic field, they may become sufficiently large, resulting in a continuous quantum phase transition at $T=0$.
The behaviors of physical quantities around such quantum critical point (QCP) have been intensively studied experimentally and theoretically, mainly in antiferromagnetic (AFM) systems~\cite{Stewart_RMP_01, Stewart_RMP_06, Lohneysen_RMP_07}.
In ferromagnetic (FM) metals, on the other hand, when the Curie temperature $T_{\rm C}$ approaches zero, FM transition tends to become of first order at a tricritical point (TCP)~\cite{Belitz_PRL_05} as actually observed in ZrZn$_{2}$($T_{\rm C}(P = 0) = 28.5$ K and $T_{\rm C}(1.65$ GPa$) \simeq 5.4$ K at TCP)~\cite{Uhlarz_PRL_04} and UGe$_{2}$($T\rm_{C}(0) = 52$ K and $T\rm_{C}(1.42$ GPa$) = 24$ K at TCP)~\cite{Huxley_PhysicaB_00}, where it is difficult to study the behaviors associated with quantum fluctuations.
Exceptions are YbNi$_{4}$(P$_{1-x}$As$_{x}$)$_{2}$~\cite{Steppke_Science_13}, Nb$_{1-x}$Fe$_{2+x}$~\cite{Brando_PRL_08}, and SrCo$_{2}$(Ge$_{1-x}$P$_{x}$)$_{2}$~\cite{Jia_NaturePhys_11}, in which $T_{\rm C}\to0$ tuning seems to be realized.
However, since these materials are non-cubic, anisotropies inherent both in magnetic interactions and conduction electrons would make theoretical analysis complicated. 
In this regard, materials with cubic crystal structures are desirable for the study of FM QCP.
In this paper, we report that a new cubic cage compound SmPt$_{2}$Cd$_{20}$ has a second-order FM transition with low $T_{\rm C} ( = 0.64$ K) \textcolor{black}{probably} accompanied by substantial quantum fluctuations, providing a good candidate starting material to approach a FM QCP.

SmPt$_{2}$Cd$_{20}$ is a member of the cage structure system $\it RTr\rm_{2}\it X\rm_{20}$ ($\it R\rm = $ rare earth, $\it Tr\rm = $ transition metal, and $\it X\rm =$ Al, Zn, and Cd), which crystallizes in the CeCr$_{2}$Al$_{20}$-type cubic structure (space group $Fd\bar{3}m$) with a cubic $T_{d}$ symmetry at the $\it R\rm$ site~\cite{Krypyakevych_DANU_68, Niemann_JSSC_95, Nasch_ZNB_97, Thiede_JAC_98, Burnett_JSSC_14}.
$\it RTr_{\rm2}\it X_{\rm20}$ compounds have gathered much interest in recent years because of a wide variety of strongly correlated electron behaviors~\cite{Torikachvili_PNAS_07, Sakai_JPSJ_10, Onimaru_PRL_11, Yazici_PRB_14_SmT2Cd20}.
In Sm$\it Tr\rm_{2}$Al$_{20}$ ($\it Tr\rm =$ Ti, V, Cr, and Ta), $c$-$f$ hybridization is relatively strong. This feature is reflected in unusual field-insensitive phase transitions and heavy-fermion (HF) behaviors, and $\log T$-dependent resistivity~\cite{Higashinaka_JPSJ_11_SmTi2Al20, Sakai_PRB_11, Yamada_JPSJ_13}. 
The magnetic susceptibility ($\chi$) exhibits an anomalous weak temperature dependent behavior with a local minimum at around 50 K for $\it Tr\rm =$ Ti and 150 K for $\it Tr\rm =$ Ta~\cite{Higashinaka_JPSJ_11_SmTi2Al20, Yamada_JPSJ_13}.
The Sm ions in these compounds are in a mixed valence state with an average Sm ion valence of about 2.85~\cite{Higashinaka_JPSConf_14, Yamada_JPCS_16}.
On the contrary, Sm$\it Tr\rm_{2}$Zn$_{20}$ ($\it Tr\rm =$ Fe, Ru, Os, Co, Rh, and Ir) and Sm$\it Tr\rm_{2}$Cd$_{20}$ ($\it Tr\rm =$ Ni and Pd) show rather localized $4\it f$ electron states with weak $c$-$f$ hybridization.
This feature is inferred from the clear Curie-Weiss behavior in $\chi(T)$ at low temperatures and the absence of the Kondo scattering ($-\log T$ dependence) in the electrical resistivity~\cite{Jia_PRB_09, Taga_JPSJSB_12, Yazici_PRB_14_SmT2Cd20, Burnett_JSSC_14, Isikawa_JPSJ_14, Isikawa_JPSJ_16, Tanahashi_JPSConf_16}.
Many of the Zn- and Cd-based compounds have magnetic ground states. So far, four Sm-based ferromagnets have been reported, i.e., SmFe$_{2}$Zn$_{20}$ ($T_{\rm C} =$ 47.4 K), SmRu$_{2}$Zn$_{20}$ ($T_{\rm C} =$ 7.6 K), SmOs$_{2}$Zn$_{20}$ ($T_{\rm C} =$ 3 K), and SmNi$_{2}$Cd$_{20}$ ($T_{\rm C} =$ 7.5 K)~\cite{Yazici_PRB_14_SmT2Cd20, Isikawa_JPSJ_14, Tanahashi_JPSConf_16}. As demonstrated in this paper, $T_{\rm C} = 0.64$ K of the new member SmPt$_{2}$Cd$_{20}$ is the lowest among Sm$\it Tr\rm_{2}\it X\rm_{20}$ compounds.

\begin{table}[b]
\caption{\label{t1}
Atomic coordinates and thermal parameters of SmPt$_2$Cd$_{20}$ at room temperature determined by single-crystal X-ray measurements. $R$ and $wR$ are reliability factors and $B_{\rm eq}$ is the equivalent isotropic atomic displacement parameter. Standard deviations in the positions of the least significant digits are given in parentheses.
}
\begin{center}
\begin{tabular}{cccccc}
\hline
\multicolumn{2}{c}{$Fd\bar3m$ ($\sharp$227)} & \multicolumn{2}{c}{$a$ $=$ 15.6237(15) $\rm\AA$ } & \multicolumn{2}{c}{$V$ $= $ 3813.7(6) $\rm\AA^3$}\\
\multicolumn{2}{c}{(origin choice 2)}  &\multicolumn{3}{c}{Position}\\
\cline{3-5}
Atom & site & $x$ & $y$ & $z$ & $B_{\rm eq}$($\rm\AA^2$)\\
\hline
Sm & $8a$     ($\bar43m$) & 1/8 & 1/8 & 1/8 & 0.69(3) \\
Pt & $16d$     ($.\bar3m$) & 1/2 & 1/2 & 1/2 & 0.76(2) \\
\textcolor{black}{Cd(1)} & $96g$     ($..m$) & 0.06051(4) & 0.06051(4) & 0.32254(5) & 1.35(2) \\
Cd(2) & $48f$($2.mm$) & 0.48716(7) & 1/8 & 1/8 & 0.95(2) \\
\textcolor{black}{Cd(3)} & $16c$     ($.\bar3m$) & 0 & 0 & 0 & 1.73(4) \\
\hline
\multicolumn{3}{c}{$R$ $=$ 2.61$\%$, $wR$ $=$ 5.43$\%$}\\
\end{tabular}
\end{center}
\end{table}

Single crystals of SmPt$_2$Cd$_{20}$ were prepared by Cd self-flux method. Chips of Sm (Furuuchi $99.9\%$), powders of Pt (Tanaka Kikinzoku $99.95\%$), and grains of Cd (Hikotaro Shudzui) were placed in an alumina crucible with an atomic ratio of 1:2:40 for Sm:Pt:Cd, and sealed in an evacuated quartz tube. The sealed tube was heated up to 900 $^{\circ}$C, kept for 5 hours, cooled to 650 $^{\circ}$C, then slowly cooled to 500 $^{\circ}$C for 75 hours ($-2^{\circ}$C$/$h). At 500 $^{\circ}$C, the tube was centrifuged to remove the excess Cd flux. 
\textcolor{black}{
Typical size of obtained single crystals is approximately \textcolor{black}{1$\times$1$\times$2mm$^{3}$}.} 
\textcolor{black}{For sample quality evaluation, we have performed elemental analysis using an X-ray fluorescence spectrometer JSX-1000S (JEOL). No impurity elements have been detected.}
Single crystal structural analysis was performed using a Rigaku XtaLABmini with graphite monochromated Mo-K$\alpha$ radiation.
\textcolor{black}{The structural parameters at room temperature, refined using the program SHELX-97~\cite{SHELX-97}, are summarized in Table~\ref{t1}.}
The lattice parameter $a = 15.6237(15) \rm\AA$ was found to be the largest among the Sm$\it Tr\rm_{2}\it X\rm_{20}$ family~\cite{Kangas_JSSC_12, Yazici_PRB_14_SmT2Cd20}. 
\textcolor{black}{The large atomic displacement parameter $B_{\rm eq}$ of \textcolor{black}{Cd(3)} at the $16c$ site is a common feature in $RTr_{2}X_{20}$~\cite{Nasch_ZNB_97, Kangas_JSSC_12, Yamada_JPSJ_13, Burnett_JSSC_14}. This finding suggests low-frequency vibrations of \textcolor{black}{Cd(3)} ions located in a large CN 14 polyhedron~\cite{Safarik_PRB_12, Hasegawa_JPCS_12, Koza_PCCP_14}.}
Note that $B_{\rm eq}$ of \textcolor{black}{Cd(1)} at the $96g$ site also has a large value in SmPt$_2$Cd$_{20}$, in contrast with normal values in $\it R\rm$Ni$_{2}$Cd$_{20}$ and $\it R\rm$Pd$_{2}$Cd$_{20}$~\cite{Burnett_JSSC_14}.

Electrical resistivity $\rho$ was measured using a standard AC four-probe technique with a physical property measurement system [PPMS; Quantum Design (QD)] combined with a homemade adiabatic demagnetization refrigerator down to 0.27 K. DC magnetization measurement was carried out using a magnetic property measurement system (MPMS; QD) down to 2.0 K and up to 7 T. Specific heat measurements were performed using a quasi-adiabatic method with the QD PPMS and a dilution refrigerator down to 0.25 K and up to 9 T.
\textcolor{black}{For these measurements, single crystals were oriented by Laue X-ray method.}


\begin{figure}[b]
\centering
\includegraphics[width=\linewidth]{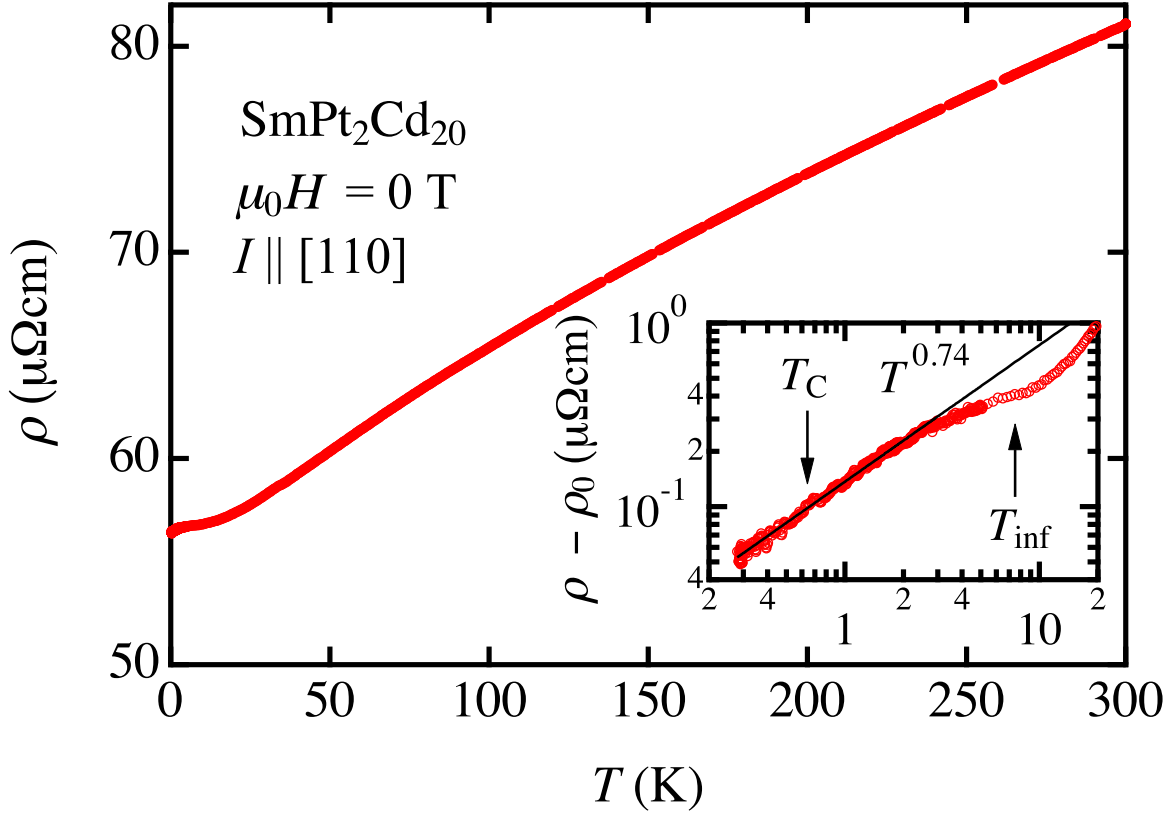}
\caption{\label{f1} Temperature dependence of resistivity $\rho$ for single-crystalline SmPt$_{2}$Cd$_{20}$ in 0 T. Inset shows the enlarged view of the temperature dependence of $\rho$ below 20 K. The temperature $T_{\rm inf} (= 7.5$ \textcolor{black}{K}) is defined using $d^2 \rho(T)/d T^2|_{T_{\rm inf}}=0$. $\rho(T)$ shows a power-law behavior with \textcolor{black}{$T^{0.74}$} below 2 K without showing any noticeable anomaly at $T_{\rm C}$.}
\end{figure}

\begin{figure}[b]
\centering
\includegraphics[width=\linewidth]{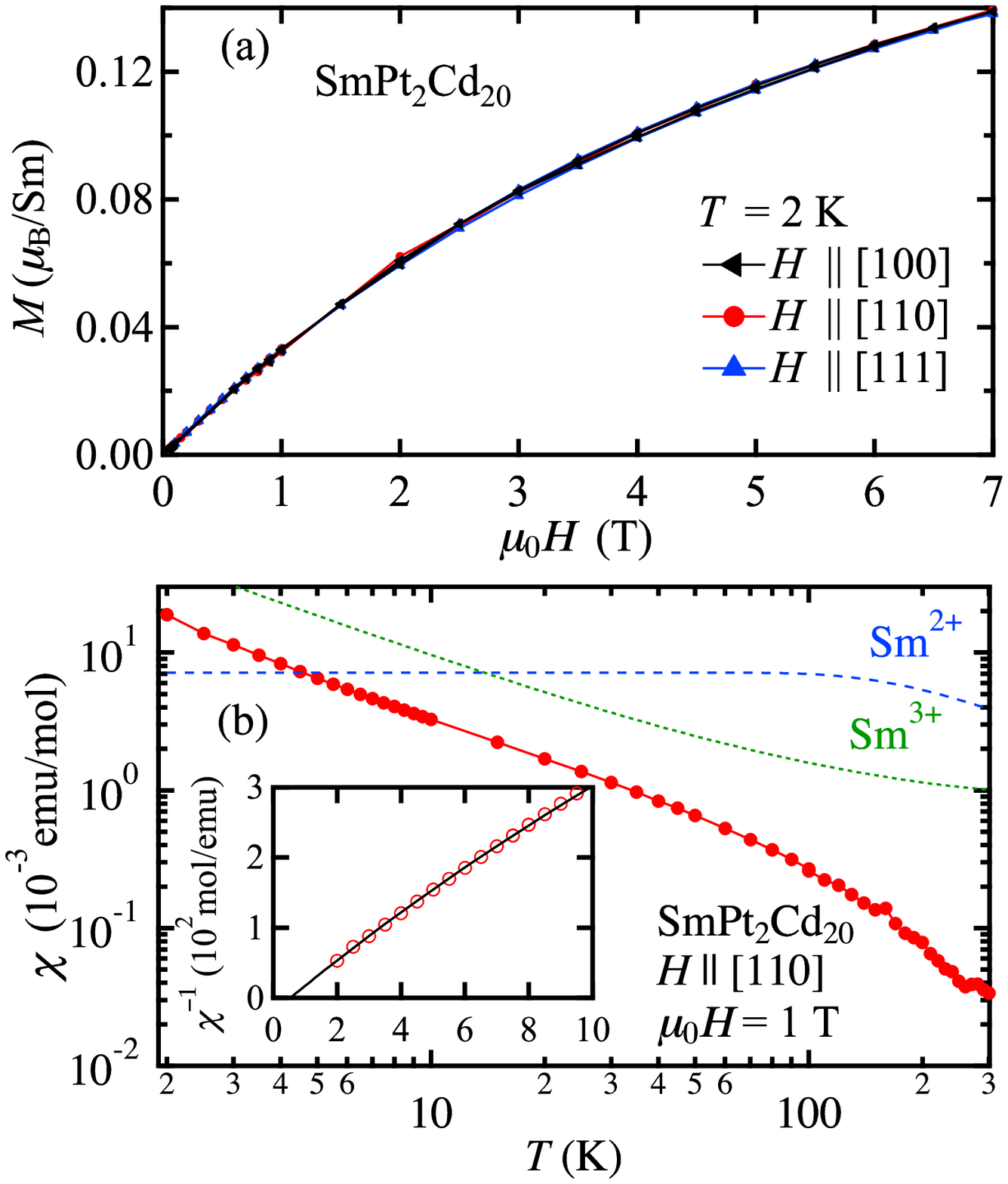}
\caption{\label{f2} (a) Field dependence of magnetization at 2 K. (b) Temperature dependence of $\chi$ in 1 T. The dotted and the dashed curves are the calculated magnetic susceptibility for free Sm$^{3+}$ and Sm$^{2+}$ ions, respectively\textcolor{black}{~\cite{MV}}. 
Inset: temperature dependence of inverse magnetic susceptibility $\chi^{-1}$. The solid curve represent a Curie-Weiss fitting between 2 and 10 K.}
\end{figure}

The temperature dependence of resistivity $\rho(T)$ in zero field depicted in Fig.~\ref{f1} shows a metallic behavior.
At around 10 K, $\rho(T)$ shows a plateau-like structure, which is characterized by the temperature $T_{\rm inf}$ (= 7.5 K) defined by $d^2 \rho(T)/d T^2|_{T_{\rm inf}}=0$.
As shown in the inset of Fig.~\ref{f1}, $\rho(T)$ shows a power-law behavior below 2 K, which can be expressed by $\rho = \rho_{0} + AT^{n}$ with $\rho_{0} = 56.4 \mu\Omega$cm, $n=0.74 \pm 0.01$, and $A = 1.37\times10^{-7} \mu\Omega$cm$/{\rm K}^n$\textcolor{black}{~\cite{rho0}}. 


\begin{figure}[b]
\centering
\includegraphics[width=\linewidth]{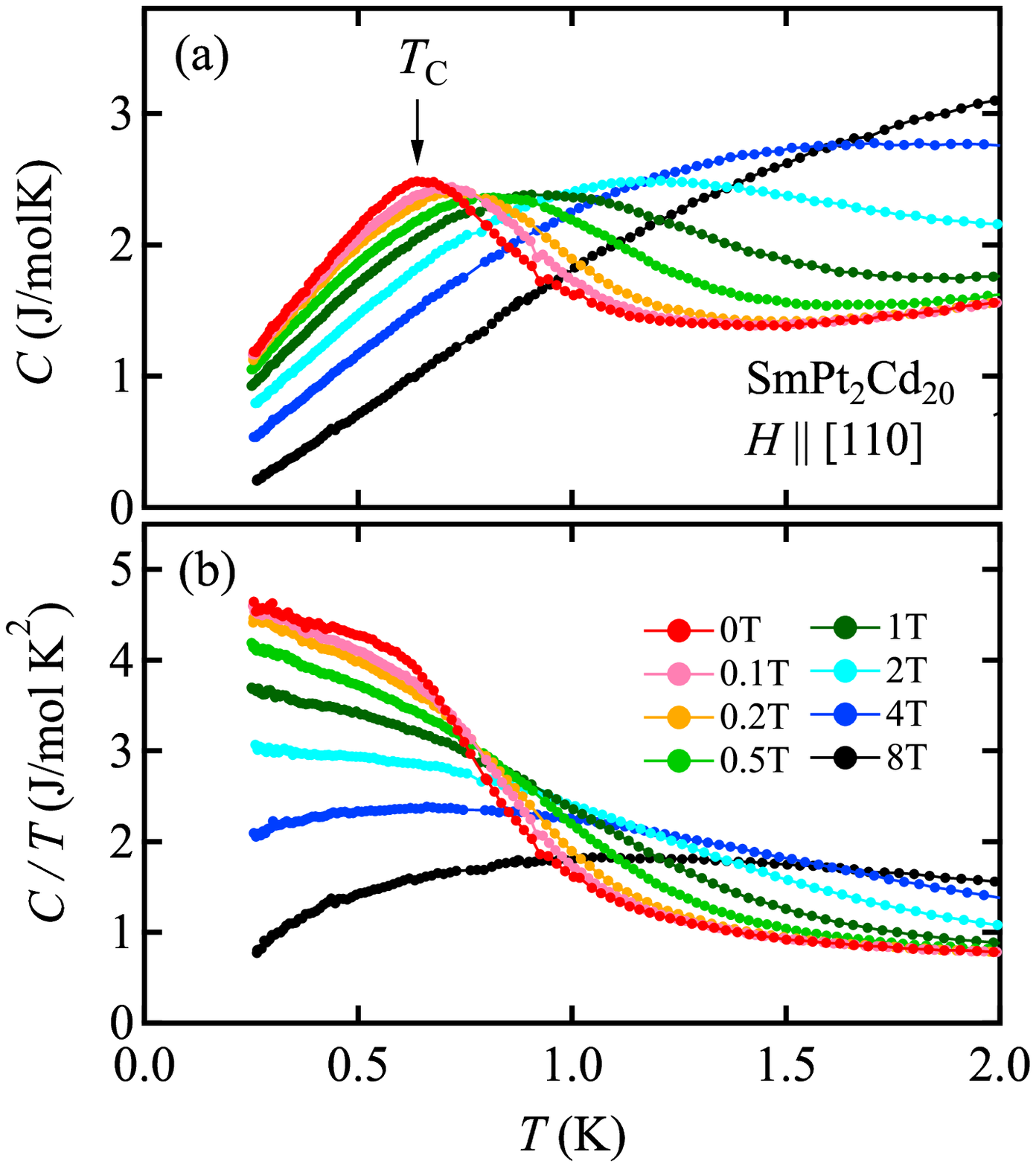}
\caption{\label{f3} Temperature dependence of (a) $C$ and (b) $C/T$. The arrow indicates the ferromagnetic transition temperature $T_{\rm C}$.}
\end{figure}

Figure~\ref{f2}(a) shows the field dependence of magnetization at 2 K for the three principal axes [100], [110], and [111]. Within the experimental uncertainty, magnetic anisotropy was not observed. Figure~\ref{f2}(b) shows temperature dependence of magnetic susceptibility $\chi ( = M/H )$ measured in 1 T along [110].   The $\chi$ data increases monotonically with decreasing temperature, demonstrating that Sm ions have a magnetic moment.
From a Curie-Weiss fitting between 2 and 10 K using
\begin{equation}
\chi = \chi_{0} + \frac{N_{\rm A}\mu_{\rm eff}^{2}}{3k_{\rm B}(T-\theta_{\rm CW})},
\label{e1}
\end{equation}
where $N_{\rm A}$ is Avogadro's constant and $k_{\rm B}$ is the Boltzmann constant, $\chi_{0} = 0.48\times10^{-3}$ emu/mol, a Curie-Weiss temperature of $\theta_{\rm CW} = 0.53$ K and an effective magnetic moment $\mu_{\rm eff} = 0.46 \mu_{\rm B}/$Sm are obtained. 
The fit result is reproduced in the inset of Fig.~\ref{f2}.
The positive value of $\theta_{\rm CW}$ indicates the existence of ferromagnetic interactions between Sm ions. 
The value of $\mu_{\rm eff}$ is smaller than $0.845 \mu_{\rm B}/$Sm for a free Sm$^{3+}$ ion. This suppression is attributable to the CEF effect. In the cubic symmetry, the $J = 5/2$ multiplet of a Sm$^{3+}$ ion splits into a $\Gamma_{7}$ doublet and a $\Gamma_{8}$ quartet. The value of $\mu_{\rm eff}$ is closer to $0.412 \mu_{\rm B}/$Sm expected for a $\Gamma_{7}$ doublet state than $0.665 \mu_{\rm B}/$Sm for a $\Gamma_{8}$ quartet state.
From $\mu_{\rm eff} \equiv g^* \sqrt{s(s+1)}\mu_{\rm B} = 0.46 \mu_{\rm B}/$Sm with an effective spin $s=1/2$ for the $\Gamma_{7}$ doublet, the effective $g$-value $g^{*}=0.53$ is obtained.
At high temperatures, $\chi$ is suppressed below the Curie-Weiss curve. 
This is mainly attributable to ion-core diamagnetic contributions from Pt and Cd ions, the total of which is estimated to be $-0.496\times10^{-3}$ emu/mol-f.u. using $-28 \times 10^{-6}$ emu/mol-Pt and $-22 \times 10^{-6}$ emu/mol-Cd ions\textcolor{black}{~\cite{ion-core, background}.}

\begin{figure}[b]
\centering
\includegraphics[width=0.9\linewidth]{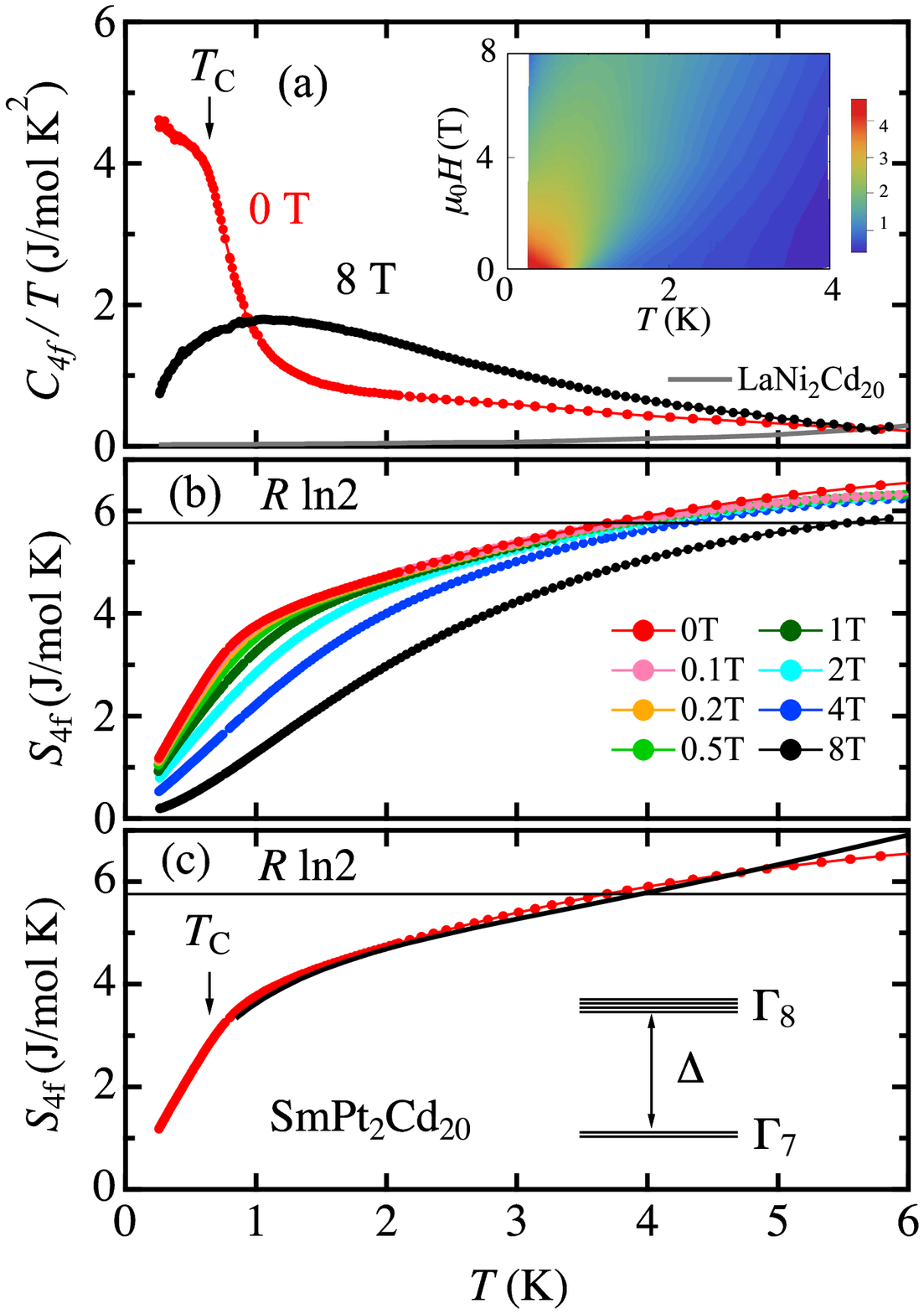}
\caption{\label{f4}(a) Temperature dependence of $C_{4\it f}/T$. The $C/T$ data for LaNi$_{2}$Cd$_{20}$ taken from Ref.~\cite{Burnett_JSSC_14} are shown for comparison. Inset: counter plot of $C_{4\it f}/T$ in the field-temperature plane. (b) Temperature dependence of magnetic entropy at selected fields between 0 and 8 T. (c) The comparison of magnetic entropy with a single-site Kondo model in zero field (see text for details).}
\end{figure}


The temperature dependence of specific heat $C(T)$ in selected magnetic fields up to 8 T is shown in Fig.~\ref{f3}(a). In zero field, $C(T)$ exhibits a slightly broadened peak at 0.64 K, indicating the appearance of a second-order phase transition\textcolor{black}{~\cite{order_of_FM}}. The peak temperature being close to $\theta_{\rm CW} = 0.53$ K suggests that the phase transition is of a FM type.
We define the Curie temperature $T_{\rm C}$ as the peak temperature in $C(T)$. \textcolor{black}{In applied fields, the peak structure becomes broader and shifts to higher temperatures as expected for a FM transition\textcolor{black}{; a thermodynamic analysis of the specific heat data shows the development of FM spontaneous magnetization~\cite{Supplementary}}.
The temperature dependence of $C$ divided by $T$ is shown in Fig.~\ref{f3}(b). In zero field, $C/T$ continues to increase anomalously below $T_{\rm C}$ with decreasing temperature, which is distinctively different from ordinary classical FM transitions, where $C/T$ decreases below $T_{\rm C}$.
This anomalous enhancement in $C/T$ approaching $T \to 0$ indicates the existence of substantial quantum magnetic fluctuations.
}

For a rough estimation of the $4\it f$ electron contribution $C_{4\it f}$ at low temperatures, we tentatively assume $C = C_{\rm 4\it f} + C_{\rm el}+ C_{\rm ph}$, where $C_{\rm el}$ and $C_{\rm ph}$ represent the conduction-electron and phonon contributions, respectively~\cite{Cnuc}.
For $C_{\rm el}+C_{\rm ph}$, the data of a nonmagnetic reference compound LaNi$_{2}$Cd$_{20}$ are used ($\simeq \gamma T+\beta T^{3}$ with $\gamma = 25.9$ mJ/mol K$^{2}$ and $\beta = 4.56$ mJ/mol K$^{4}$ below 2 K)~\cite{Yazici_PRB_14_SmT2Cd20}.
The extracted $C_{4\it f}/T$ and the magnetic contribution to the entropy $S_{\rm4\it f}(T) = \int_0^T(C_{4\it f}/T')dT'$ are shown in Fig.~\ref{f4}(a, b).
All the $S_{\rm4\it f}(T)$ curves for $\mu_0 H \le 8$ T show a plateau
behavior at around $R\ln2$, suggesting that the CEF ground state is a
$\Gamma_{7}$ doublet; contrarily, Sm$\it Tr\rm_{2}$Al$_{20}$ tends to have a $\Gamma_{8}$ quartet ground state~\cite{Higashinaka_JPSJ_11_SmTi2Al20, Sakai_PRB_11, Yamada_JPSJ_13}.
A $\Gamma_{7}$ doublet carries only magnetic dipole moments and no
higher-rank multipoles as active degrees of freedom~\cite{Shiina_JPSJ_97, Santini_RMP_09}.
This feature makes SmPt$_{2}$Cd$_{20}$ as a suitable candidate system to study FM QCP.

In zero field, $S_{\rm4\it f}$ is largely suppressed above $T_{\rm C}$ and $S_{\rm4\it f}(T_{\rm C})$ is only $0.5\times R\ln2$.
This suppression may be caused by Kondo effect and/or FM fluctuations (short-range ordering) in $T > T_{\rm C}$.
Tentatively, $S_{4\it f}(T)$ above $T_{\rm C}$ is compared with a Kondo model for a CEF-split $J = 5/2$ ion in the Kondo regime as shown in Fig.~\ref{f4}(c)~\cite{Kuramoto_ZPB_83, Bickers_RMP_87, CXcal-excel}.
The best fit to the experimental data yields the Kondo temperature $T_{\rm K}^{(6)} = 6$ K (for the $J = 5/2$ sextet) and the CEF splitting $\Delta = 30$ K with a $\Gamma_{7}$ ground state.
Using $T_{\rm K}^{(6)3}=\Delta^{2} T_{\rm K}^{(2)}$ ~\cite{Yamada_PTP_84}, the Kondo temperature of the ground state doublet is estimated to be $T_{\rm K}^{(2)} = 0.24$ K. 
The fact that $T_{\rm C}$ and $T_{\rm K}^{(2)}$ are similar order may point to the competition between Kondo effect and the FM interactions in SmPt$_{2}$Cd$_{20}$.
\textcolor{black}{Considering $\Delta = 30$ K, the plateau behavior in $\rho(T)$ at around $T \sim T_{\rm inf}=7.5$ K (see the inset of Fig.~\ref{f1}) can be understood as a crossover; in $T >T_{\rm inf}$, inelastic conduction-electron scattering associated with the $\Gamma_{7}$ - $\Gamma_{8}$ CEF levels decreases with lowering temperature and, in $T <T_{\rm inf}$, anomalous scattering with the $T^{0.74}$ behavior develops with lowering temperature.}


The contour plot of $C_{4\it f}/T$ shown in the inset of Fig.~\ref{f4}(a) shows clearly that the magnetic entropy is released significantly in a low-$T$ and low-$H$ region.
This region has a broad tail extending into higher fields, which corresponds to a broad maximum in the $C_{4\it f}/T$ vs $T$ curve.
In 8 T, the maximum temperature is 1.1 K, which is slightly higher than 0.88 K of a Schottky peak maximum caused by the Zeeman splitting of the $\Gamma_7$ doublet ($g^* \mu_{\rm B} H/k_{\rm B}=2.85$ K), probably due to FM interactions among Sm ions.

As the $R \ln2$ plateau behavior in $S_{4\it f}(T)$ shows, the $4\it f$ electrons in SmPt$_{2}$Cd$_{20}$ have a rather localized nature in this temperature range.
This is in good agreement with the similar trend observed in Sm$\it Tr\rm_{2}\it X\rm_{20}$ with $\it X\rm =$ Zn and Cd~\cite{Taga_JPSJSB_12, Isikawa_JPSJ_14, Yazici_PRB_14_SmT2Cd20, Isikawa_JPSJ_16}, in marked contrast with strongly-hybridized characters in $\it X\rm = Al$ compounds, e.g., mixed valent Sm ion state and $-\log T$ behavior in $\rho(T)$~\cite{Higashinaka_JPSJ_11_SmTi2Al20, Sakai_PRB_11, Yamada_JPSJ_13, Higashinaka_JPSConf_14, Yamada_JPCS_16}.


At FM QCP in metals, the theoretically expected behaviors of physical quantities are $\rho \sim T^{5/3}$, $\chi^{-1} \sim T^{4/3}$ and $C/T \sim -\ln T$~\cite{Hertz_PRB_76, Millis_PRB_93, Moriya_Book_85}.
These are clearly different from the observed unconventional behaviors in SmPt$_{2}$Cd$_{20}$, i.e., $\rho \sim T^{0.74}$~\cite{Ein_temp,Moriya_JPSJ_75} and increasing $C_{4\it f}/T$ with $T \to 0$.

This discrepancy may be due to a finite deviation from a FM QCP in some control parameters (i.e., the non-zero $T_{\rm C}$).
\textcolor{black}{Note that} theoretical considerations indicate that the quantum critical regime can also extend into a magnetically ordered phase and singular behaviors can appear below $T_{\rm C}$~\cite{Lohneysen_RMP_07}, which may correspond to the present observation.

For the $T_{\rm C} \to 0$ tuning in SmPt$_{2}$Cd$_{20}$, applying hydrostatic pressure is not a likely tool since pressure generally stabilizes Sm$^{3+}$ relative to Sm$^{2+}$ because of the larger ion radius of Sm$^{2+}$ and, in SmPt$_{2}$Cd$_{20}$, Sm$^{3+}$ is already attained as $\chi(T)$ shows.
On the other hand, La doping for Sm will be a promising means.
In SmTi$_{2}$Al$_{20}$, it has been demonstrated that this doping actually decreases the magnetic ordering temperature~\cite{Higashinaka_JPCS_16}.
Note that each Sm (La) ion is located at the center of the cubic cage structure formed by 16 $X$ ions and separated each other in the crystal structure. This feature of the cage compounds should help to minimize randomness effects that is inevitable in doping experiment, as proved by NQR local probe in filled skutterudites~\cite{Yogi_JPSJ_06}. It would be highly interesting to investigate how the physical quantities behave at a metallic FM QCP\textcolor{black}{, when it is attained,} in cubic (Sm$_{1-x}$La$_x$)Pt$_{2}$Cd$_{20}$.

This work was supported by MEXT/JSPS KAKENHI Grant Numbers 15J07600, 24740239, 15H03693, 15K05178, 23540421, 23340107, and 20102005.

\nocite{*}

\bibliography{apssamp}

\begin{thebibliography}{9}
\bibitem{CXcal-excel} CXcal-excel; a free program coded by O. Sakai. It provides the NCA calculation of specific heat and magnetic susceptibility of a $J = 5/2$ single ion with a cubic symmetry in the Kondo regime.
\end{thebibliography}

\end{document}